# The Land Surface Temperature Synergistic Processor in BEAM: A Prototype towards Sentinel-3

Ana Belen Ruescas *, Olaf Danne, Norman Fomferra and Carsten Brockmann

Brockmann Consult GmbH, Max-Plack Str.2, 21502 Geesthacht, Germany;
olaf.danne@brockmann-consult.de (O.D.); norman.fomferra@brockmann-consult.de (N.F.);
casrten.brockmann@brockmann-consult.de (C.B.)
* Correspondence: ana.ruescas@brockmann-consult.de; Tel.: +49-415-288-9300

**Abstract:** Land Surface Temperature (LST) is one of the key parameters in the physics of land-surface processes on regional and global scales, combining the results of all surface-atmosphere interactions and energy fluxes between the surface and the atmosphere. With the advent of the European Space Agency (ESA) Sentinel 3 (S3) satellite, accurate LST retrieval methodologies are being developed by exploiting the synergy between the Ocean and Land Colour Instrument (OLCI) and the Sea and Land Surface Temperature Radiometer (SLSTR). In this paper we explain the implementation in the Basic ENVISAT Toolbox for (A)ATSR and MERIS (BEAM) and the use of one LST algorithm developed in the framework of the Synergistic Use of The Sentinel Missions For Estimating And Monitoring Land Surface Temperature (SEN4LST) project. The LST algorithm is based on the split-window technique with an explicit dependence on the surface emissivity. Performance of the methodology is assessed by using MEdium Resolution Imaging Spectrometer/Advanced Along-Track Scanning Radiometer (MERIS/AATSR) pairs, instruments with similar characteristics than OLCI/ SLSTR, respectively. The LST retrievals were properly validated against *in situ* data measured along one year (2011) in three test sites, and inter-compared to the standard AATSR level-2 product with satisfactory results. The algorithm is implemented in BEAM using as a basis the MERIS/AATSR Synergy Toolbox. Specific details about the processor validation can be found in the validation report of the SEN4LST project.

**Keywords:** sentinels; sensor synergy; OLCI; SLSTR; land surface temperature; BEAM

## 1. Introduction

The Sentinel satellite constellation series is developed by the European Space Agency (ESA) in order to support European operational services and the policy needs of the Copernicus programme. The first three Sentinel missions contribute to the understanding of the Earth System by detecting, monitoring and assessing changes in ocean, cryosphere, and land components [1,2]. In particular, the Sentinel 3 (S3) mission provides continuity to Environmental Satellite (ENVISAT) capabilities while contributing to a number of services related to ocean and land products [3]. Main instruments on board the S3 mission are the Ocean and Land Colour Imager (OLCI) and the Sea and Land Surface Temperature Radiometer (SLSTR). OLCI is a push-broom imaging spectrometer instrument building on the heritage of ENVISAT MERIS (MEdium Resolution Imaging Spectrometer), with 21 spectral bands covering the 0.4–1 µm range at maximum 300 m spatial resolution [4]. SLSTR is a dual view conical imaging radiometer building on the heritage of ENVISAT AATSR (Advanced Along-Track Scanning Radiometer). It includes 9 spectral bands covering the 0.5–12 µm spectral range, with two additional bands in the near infrared for clouds and aerosols retrieval and other two for active fire detection. The SLSTR spatial resolution is 500 m for the visible and near-infrared (VNIR) bands and 1km for the thermal infrared (TIR) and fire bands [5], while the swath is 1400 km for the single view and 740 km for the double view observation. One of the main objectives of the S3 mission is to provide Europe with continuity of the ENVISAT type measurement capability to determine sea, ice and land surface temperature. In this context, ESA funded the project "Synergistic Use of The Sentinel Missions For Estimating And Monitoring Land Surface Temperature (SEN4LST)" [6,7], which had the main objective to fully utilize the synergy between SLSTR and OLCI instruments to improve atmospheric correction (including cloud screening) and land surface emissivity (LSE) characterization for a better estimation of the land surface temperature (LST). As a result of the SEN4LST project, a new land surface temperature processor plug-in has been implemented in the Basic ENVISAT Toolbox for (A)ATSR and MERIS (BEAM). The synergistic algorithm in which is based is designed and implemented for improving cloud screening, global aerosol and atmospheric correction using the combined multi-spectral and multi-angle information from instrument pair measurements [8]. The implemented processor, the functionality of which is the main target of this paper, supports both MERIS/AATSR and OLCI/SLSTR (Sentinel 3) data pairs. Because there are still not available pairs of the OLCI/SLSTR, simulated data is used instead [9].

## 2. The BEAM Toolbox and the Land Surface Temperature Processor

The main target of this paper is to present to the scientific community the Land Surface Temperature BEAM processor. This plug-in will offer interested researchers the possibility to generate an LST product with MERIS and AATSR images (and future OLCI/SLSTR), comparable in quality to the standard ESA's Level 2 LST product. The LST algorithm used in the processor has been tested and published by scientific teams; therefore, there is not an extensive explanation of the theory and physics of the algorithm or its validation, but necessary references have been cited for the convenience of the audience of the present work.

*2.1. Basic ENVISAT Toolbox for (A)ATSR and MERIS (BEAM)*

The BEAM Earth Observation Toolbox and Development Platform [10] is a collection of executable tools and Application Programming Interfaces (APIs) that have been developed to facilitate the utilisation, viewing and processing of a variety of remotely sensed data. The purpose of the BEAM is not to duplicate existing commercial packages, but to complement them with functions dedicated to the handling of data products of Earth observing satellites, specifically focused on ocean colour applications, but also used for other land surface processings. The main components of the BEAM are:

- VISAT: an intuitive desktop application used for Earth Observation (EO) data visualisation, analysis and processing (Figure 1) (version 5, Brockmann Consult GmbH, Geesthacht, Germany).
- A set of scientific data processors running either from the command-line or invoked by VISAT.
- The command-line tool *gpt* (graph processing tool) is used to execute processing graphs made up of operators nodes developed using the BEAM Graph Processing Framework (GPF, see Java API below) (Brockmann Consult GmbH, Geesthacht, Germany).
- A data product converter tool *pconvert* (Brockmann Consult GmbH, Geesthacht, Germany) allowing a user to convert raw data products to the BEAM-DIMAP standard format, to GeoTIFF, to HDF-5 or to RGB images.
- A Java$^{TM}$ API which provides ready-to-use components for remote sensing related application development and plug-in points for new BEAM extension modules. Besides a number of extension points such as product reader and writers, the BEAM API comprises the GPF, which is used to rapidly create raster data processors. The VISAT Rich Client Platform is used to develop rich Graphical User Interface (GUI) applications based on BEAM VISAT.

BEAM is programmed in pure Java$^{TM}$ to allow a maximum portability. The BEAM software (version 5, Brockmann Consult GmbH, Geesthacht, Germany) has been successfully tested under MS Windows$^{TM}$ XP®, Windows$^{TM}$ Vista, Windows$^{TM}$ 7 and also Mac OS X and as well as under Linux ("Linux" trademark is owned by Linus Torvalds and administered by the Linux Mark Institute) and Solaris® (Sun Microsystems, ORACLE, San Francisco, USA) operating systems. Since end of 2015 BEAM is not further developed. It will remain under maintenance until at least end of 2016. BEAM users are encourage using its evolutionary successor SNAP from now on (version 4, ESA). When installing SNAP along with the Sentinel-3 Toolbox (version 4, ESA) the users will have the same experience as with BEAM and they can even extend the number of features by installing other toolboxes like those for Sentinel-1 or Sentinel-2 [11].

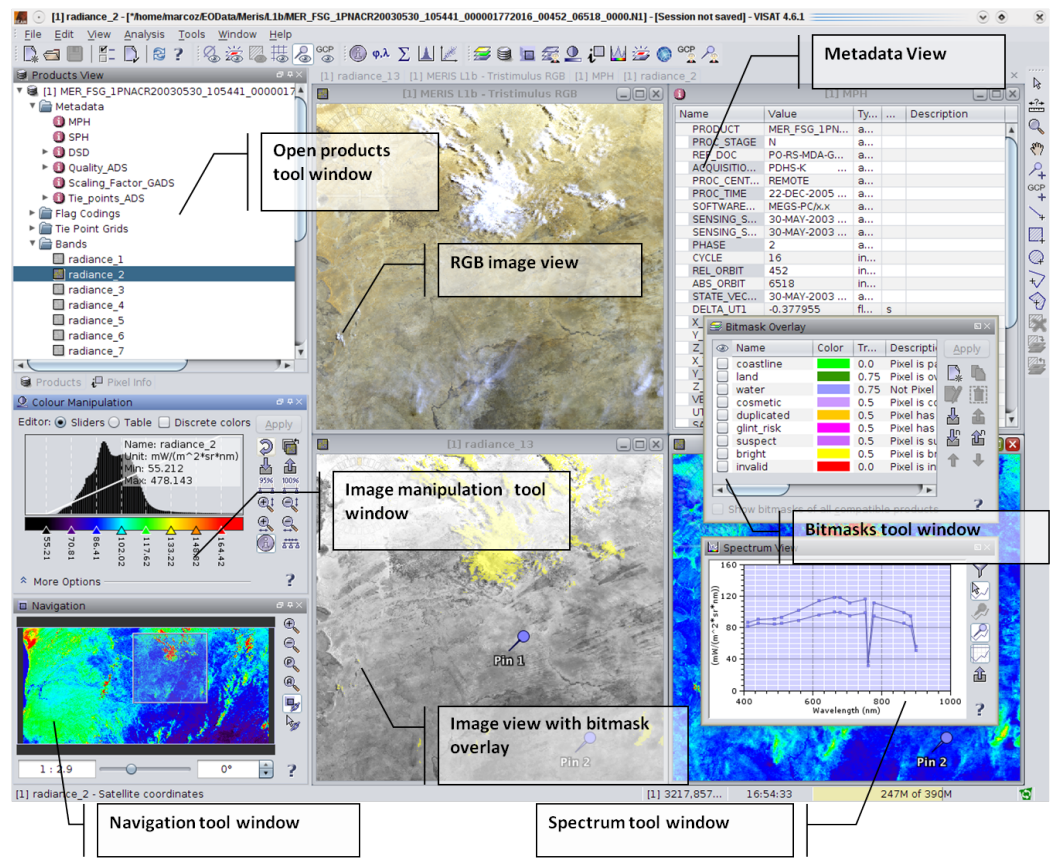

**Figure 1.** BEAM-VISAT overview.

*2.2. The LST Processor as a BEAM Plug-In*

The LST processor is a BEAM plug-in that allows the processing of MERIS/AATSR L1b pairs, and simulated OLCI/SLSTR surface directional reflectance. The final objective of this plug-in is to provide an LST product using the algorithms developed in the framework of the SEN4LST project to be applied to MERIS/AATSR and potentially the forthcoming S3 OLCI/SLSTR instruments. Inputs needed for application of the SEN4LST algorithms (e.g., the split-window algorithm given by Equation (1)) are surface directional reflectances (SDRs), which for MERIS/AATSR are retrieved in a foregoing processing step applying the MERIS/AATSR synergistic approach [8]. This method is based on an atmospheric correction scheme together with a cloud-screening procedure, both developed within the frame of ESA's MERIS/AATSR Synergy Toolbox project.

2.2.1. Theoretical Background of the LST Algorithm

A large number of land surface temperature (LST) and land surface emissivity (LSE) algorithms have already been published ([12,13]). Many of these algorithms are based on the split-window (SW) or the two-channel (TC) techniques. The SW is based on the differential absorption concept [14], in which the atmospheric effect over the measured signal is corrected using the difference signal in two thermal (TIR) bands at two different wavelengths, or by one TIR band on two different view angles, the dual-angle (DA) algorithms. The two types of algorithms depend on the surface emissivity. The retrieval of the emssivity values can be done in many ways ([15–17]). The approach selected here is based on vegetation indices (VIs) [18] and on classification-based approaches [19]. To allow for near real-time retrievals and to avoid nighttime acquisitions, methods based on day/night pairs are not considered. The proposed algorithm finally selected has the expression [20]:

$$T_S = T_i + c_1(T_i - T_j) + c_2(T_i - T_j)^2 + c_0 + (c_3 + c_4 W)(1 - \varepsilon) + (c_5 + c_6 W)\Delta\varepsilon, \tag{1}$$

where $T_s$ is the LST (in K), $T_i$ and $T_j$ are at-sensor brightness temperatures (in K), $W$ is the atmospheric water vapor content (in $gcm^{-2}$), $\varepsilon$ is the mean land surface emissivity, LSE: $0.5 \cdot (\varepsilon_i + \varepsilon_j)$, and $\Delta\varepsilon$ is the LSE difference ($\varepsilon_i - \varepsilon_j$). Sub-indices "*i*" and "*j*" refer to two different thermal infrared (TIR) bands, thus leading to the SW algorithm, or to one TIR band but two different view angles (e.g., nadir "*n*" and oblique "*o*" views), thus leading to the DA algorithm. Coefficients $c_0$ to $c_6$ are obtained from statistical regressions performed over simulated data. This algorithm is physically-based, and it explicitly includes the atmospheric water vapor (*W*) content and surface emissivity (LSE). Furthermore, it can easily incorporate external sources of W and LSE. Input of the emissivity value is calculated using the normalized difference vegetation index (NDVI) thresholds method (THM) first established by [18] (Equation (2)). The threshold method is applied on each thermal band (i) because it has a spectral base:

$$\begin{aligned}NDVI < NDVIs : \varepsilon_i &= a_i + b_i \rho_{red},\\ NDVIs \leq NDVI \leq NDVIv : \varepsilon_i &= \varepsilon_{is}(1 - Pv) + \varepsilon_{iv} Pv + C_i,\\ NDVI > NDVIv : \varepsilon_i &= 0.99,\end{aligned} \tag{2}$$

where $\rho_{red}$ is the reflectance at the red band, $\varepsilon_s$ and $\varepsilon_v$ are reference values of surface emissivity for soil and vegetation, respectively, *C* is a cavity term for rough surfaces that depends on geometrical factors (and it is neglected for operational purposes or assumed constant; in the present case is assumed to be 0.005 [21]), and *Pv* is the fractional vegetation cover, which can be obtained from the scaled NDVI [22]:

$$Pv = \frac{NDVI - NDVIs}{NDVIv - NDVIs}, \tag{3}$$

where NDVI is the current value and NDVIs and NDVIv are reference values of NDVI for bare soil and fully vegetated surfaces, respectively. Values of 0.15 for NDVIs and 0.99 for NDVIv are considered representative of global conditions [23]. A simplified version of the NDVI-THM method is selected because it requires reduced computing time and allows contemporary implementation. More details ca be found in Section 2.2.2 and Equations (5) and (6). It has the advantage that whenever improved products of *W* and LSE are available, the potential LST product generated with this algorithm can be easily reprocessed to provide a new product version.

Atmospheric Correction and Cloud Screening

It is important to highlight again that the scope of this paper is not to provide a detailed description of the atmospheric correction and the cloud screening methodologies. In the SEN4LST project documentation, there is a review and adaptation of both approaches for potential application on S3. The surface reflectances retrieved from the MERIS and AATSR sensors are necessary for a correct estimation of the LSE using the NDVI-THM method introduced before [18]. The water vapour content is also a required input (see Equation (1)). The effects of aerosols and absorbing gases are modelled as well, scattering reflectance determination being a key issue for a good atmospheric correction. The great variability of the aerosol scattering induces a high uncertainty in the derivation of surface reflectance. The magnitude of the scattering increases with view angle, that is, the off-nadir view of AATSR and SLSTR—and in high latitudes the nadir view—constitutes a very important factor, especially for S3 due to their wider swath width. The method employed in SEN4LST is based in a modification of the synergy approach ([24,25]). The output is the aerosol optical depth at reference wavelength, an estimate of the aerosol model and the Angstrom coefficient, and the atmospherically corrected reflectance. Many of the ideas and approaches performed in the MERIS/AATSR Synergy project are applied here ([26,27]). For instance, the cloud screening strategy is taken from the SYNERGY and Globalbedo cloud masks, which generates a cloud probability from an ensemble of multilayer perceptron artificial neural networks using inputs from both MERIS and AATSR. Other approaches were also tested in the SEN4LST project [7], like the Bayesian approach by [28].

2.2.2. Description of the LST Processor in BEAM

The LST processor is configured in BEAM using a two-step procedure, composed of several sub-processors. First, the surface directional reflectance (SDR) are retrieved using the synergy derivation approach mentioned in Section 2.2.1. Second, the land surface temperatures are calculated applying the LST algorithms available in the software, derived from the SEN4LST project.

Surface Directional Reflectance: The Synergy Derivation Approach

Figure 2 shows a screen-shot of the SDR process to extract the reflectances used as input in the LST algorithm.

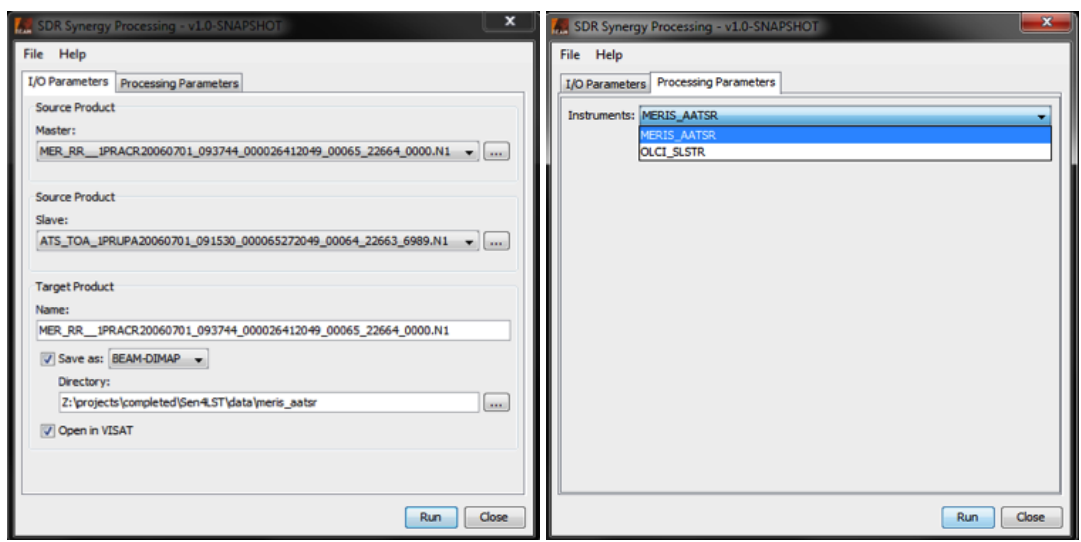

**Figure 2.** Screen-shot of the SDR module in the BEAM LST plug-in.

The first sub-process consists of the collocation of the MERIS and AATSR images (Figure 3). As input for the cloud screening and atmospheric correction steps within the Surface Directional Reflectance (SDR) retrieval, a collocated MERIS/AATSR product is generated from a MERIS L1b product and a corresponding—overlapping in time and space—AATSR L1b product. This new collocation product contains a copy of all components of the master product, i.e., band data, tie-point grids, flag coding, bitmask definitions, and metadata. The MERIS L1b is treated as the "master" product, the AATSR L1b as the "slave" product. The non-overlapping areas of the master and slave products are cropped, the band data of the slave product are then re-sampled into the geographical raster of the master product, applying a nearest-neighbour re-sampling. Consequently, the collocation algorithm requires accurate geo-positioning information for both master and slave products. Since the MERIS L1b contains radiances, but the AATSR L1b contains Top Of Atmosphere (TOA) reflectances, a radiance-to-reflectance conversion is applied to the MERIS data.

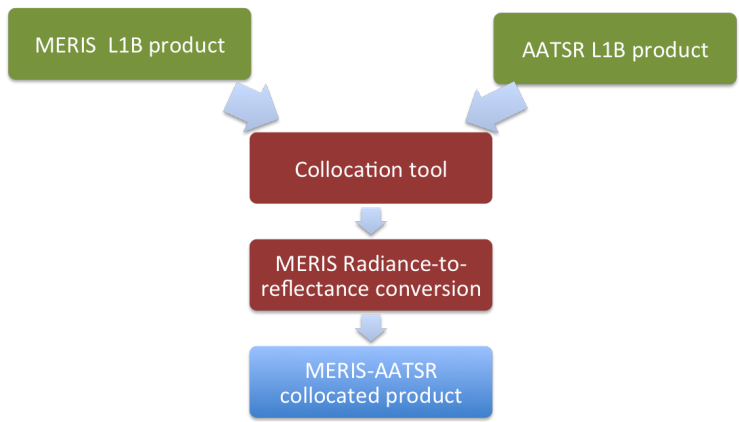

**Figure 3.** Logical flow of the L1b product collocation.

This correction is done using the MERIS Radiometric Correction Tool available in BEAM, which uses a simple equation:

$$\rho_0 = \frac{\pi L_0}{F_0 \cos(\theta_s)}, \qquad (4)$$

Table 1 shows the variables of Equation (4):

**Table 1.** Variables of Equation (4).

| Variable | Name | Unit |
| --- | --- | --- |
| $F_0$ | Solar irradiance at TOA | $Wm^{-2}nm^{-1}$ |
| $L_0$ | Radiance at TOA | $Wm^{-2}sr^{-1}nm^{-1}$ |
| $\rho_0$ | Reflectance at TOA | dl |
| $\theta_s$ | Solar zenith angle | rad |

Both radiances and converted reflectances are stored in the collocated product. The tie points are taken from the MERIS (master) product as they are, whereas the AATSR tie points are resampled onto the MERIS grid and stored as regular bands. In the SEN4LST processor realization, the collocation product serves as an intermediate result, and it is not written to disk. After the collocation follows the cloud screening, whose main objective is to accurately detect pixels contaminated by clouds and exclude these from LST retrieval. In summary, the cloud screening algorithm follows [26,27]: first, a feature extraction and selection based on meaningful physical features is carried out (clouds are bright, white, high, etc.). Then, a supervised pixel-based feature classification, based on simulated TOA radiances and corresponding cloud optical thickness from radiative transfer models (RTMs) and on the real MERIS and AATSR data from a manually labeled training set, is applied to these features providing the pixel label (cloud or cloud free) and the cloud abundance. Finally, an image-based post-processing (such as removal of coastline artifacts) is applied to the classified image to refine the cloud products. The resulting product contains all bands and tie points copied from the collocated product, and an additional flag band indicating cloudy pixels. In the SEN4LST processor realization, the cloud screening product serves as an intermediate result, and it is not written to disk.

The atmospheric correction aims at making use of the angular and spectral sampling available from MERIS and AATSR in order to develop an improved algorithm for atmospheric correction (including aerosol retrieval) over land. Basically, the problem of land aerosol and SDR retrieval represents an optimization subject to multiple constraints. Therefore, the underlying algorithm is recursive and takes as input TOA reflectance data for the solar reflective AATSR bands at both nadir and backward views (resulting in a total of eight channels), and the 14 MERIS bands at all non-absorbing channels (excluding O2 absorption band). All of these bands are available from the cloud screening output product described in the previous paragraph. SDRs for all input channels are calculated using look-up tables (LUTs). The SDRs depend on initial aerosol parameters, sun-sensor geometry, and surface pressure. This atmospheric correction retrieval scheme is illustrated in Figure 4.

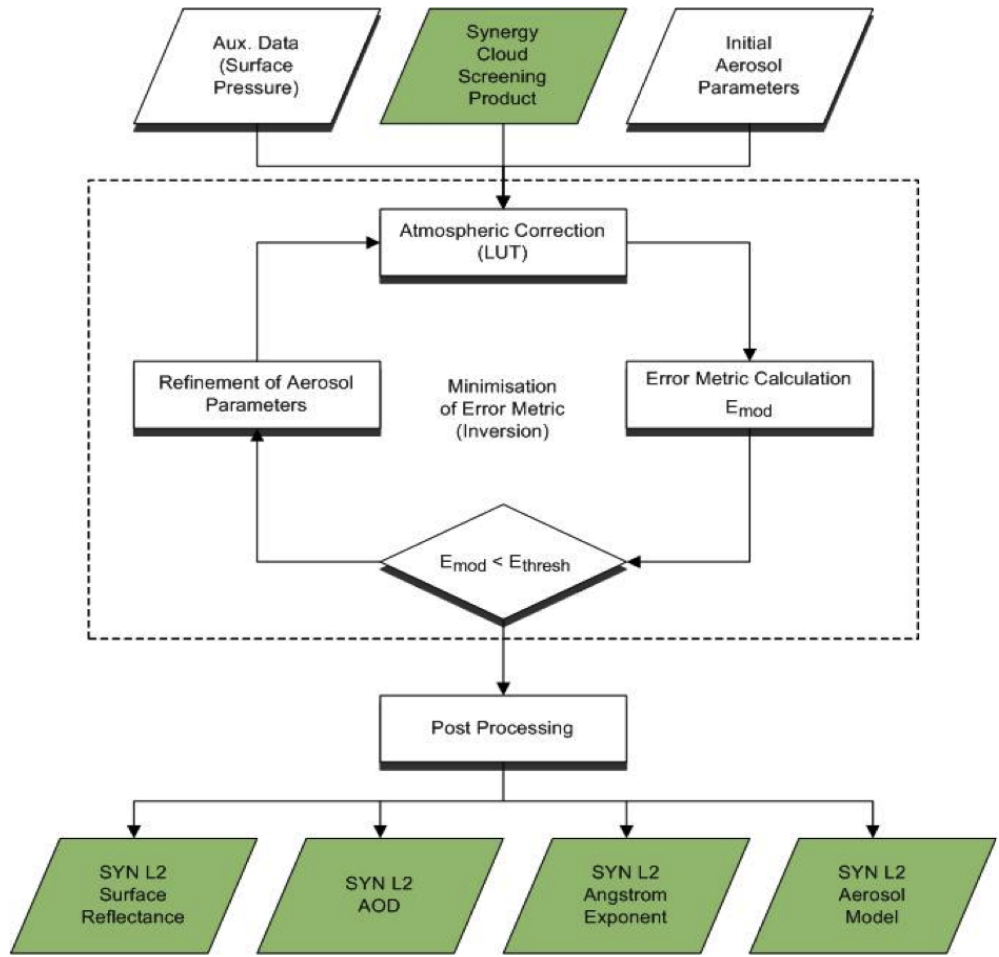

**Figure 4.** Logical flow of the retrieval of surface reflectances and aerosol properties.

The output of the algorithm are aerosol optical depth (AOD) at a reference waveband (550 nm) Angstrom exponent (an exponent that expresses the spectral dependence of aerosol optical thickness ($\tau$) with the wavelength of incident light ($\lambda$). The spectral dependence of aerosol optical thickness can be approximated (depending on size distribution) by $\tau_a = \beta \lambda^\alpha$, where $\alpha$ is Angstrom exponent ($\beta$ = aerosol optical thickness at 1 μm)), an estimate of the aerosol model, and atmospherically corrected surface reflectances (SDRs) for all bands used. Moreover, an error metric is provided where a low value of this metric corresponds to a set of surface reflectances (and hence atmospheric profile) that is realistic. Consequently, the algorithm needs to be applied recursively until an optimal aerosol/SDR solution is retrieved. However, although aerosol parameters and SDRs are available as output at the same time, the retrieval is in practice split into an aerosol and SDR part for performance reasons. As the spatial variation of aerosol in the atmosphere is generally much lower than the variation of the reflectances, the iterative retrieval scheme is first applied on a coarser grid, keeping just the optimal aerosol result. In return, this result is downscaled to the original resolution, and the algorithm is applied just one more time to derive the corresponding SDR result. For the final LST retrieval, only these SDRs will be used, and the aerosol quantities are not explicitly needed. The SDR result product contains all the flag bands, the "latitude" and "longitude" tie point grids, and the AATSR brightness temperatures copied from the input product, and as additional bands the SDRs for MERIS wavelengths 620 nm and 753 nm, and the SDRs for AATSR nadir/forward at wavelengths 555 nm and 659 nm (Table 2). In the SEN4LST processor realization, the SDR product serves as final result of the first major processor module ("SEN4LST SDR Synergy Processor"), and it is written to disk.

Table 2. SDR product dataset overview.

| Dataset | Band Name | Unit | Dimension #Bytes |
|---|---|---|---|
| All flag bands from MERIS/AATSR collocation product | l1_flags_MERIS confid_flags_nadir_AATSR confid_flags_fward_AATSR cloud_flags_nadir_AATSR cloud_flags_fward_AATSR | dl | 5*NL*NC*4 |
| Cloud flag | cloud_flags_synergy | dl | NL*NC*4 |
| Latitude and longitude tie point grids from MERIS L1 | latitude longitude | deg deg dl | 2*NL*NC*4 2*NL*NC*4 |
| AATSR nadir brightness temperature | btemp_nadir_1200_AATSR btemp_nadir_1100_AATSR btemp_nadir_0370_AATSR | K K K | 3*NL*NC*4 3*NL*NC*4 3*NL*NC*4 |
| AATSR forward brightness temperature | btemp_fward_1200_AATSR btemp_fward_1100_AATSR btemp_fward_0370_AATSR | K K K | 3*NL*NC*4 3*NL*NC*4 3*NL*NC*4 |
| MERIS surface reflectance at 620 nm and 753 nm | SynergySDR_620_MERIS SynergySDR_753_MERIS | dl dl | 2*NL*NC*4 2*NL*NC*4 |
| AATSR nadir surface reflectance at 555 nm and 659 nm | SynergySDR_nadir_555_AATSR SynergySDR_nadir_659_AATSR | dl dl | 2*NL*NC*4 2*NL*NC*4 |
| AATSR forward surface reflectance at 555 nm and 659 nm | SynergySDR_fward_555_AATSR SynergySDR_fward_AATRS | dl dl | 2*NL*NC*4 2*NL*NC*4 |

Note: confid = confidence, btemp = brightness temperature, fward = forward, SDR = surface directional reflectance, dl = dimensioless, NL = number of columns in a scan line, NC = number of scan lines in a dataset.

Land Surface Temperature Retrieval

Necessary input variables for the LST retrieval are the atmospheric water vapour, the surface emissivity, and regression coefficients (Figure 5).

The atmospheric water vapour W is obtained from the MERIS Level 2 product corresponding to the given L1b input. This product gives the total column water vapour per pixel in $gcm^{-2}$. If such a product is not available, the water vapour is set to a constant value of $W = 2.0\ gcm^{-2}$ within the retrieval.

Input surface emissivities into the LST algorithms are retrieved using the NDVI Thresholds Method described in Section 2.2.1. The MERIS/OLCI bands selected for the calculation of the NDVI are 560 nm and 665 nm. The NDVI Thresholds Method uses certain NDVI values (minimum/maximum thresholds) to identify the range between pure soil pixels (NDVImin) and pixels of full vegetation (NDVImax). For NDVImin and NDVImax fixed values might be assumed (e.g., 0.2 and 0.5, as suggested by [6]. However, the current processor uses a more dynamic approach by taking the minimum and maximum NDVI from the given scene. Then, a fractional vegetation cover (FVC or Pv) can be computed (Equation (3)) and used in the derivation of the emissivity:

$$\varepsilon_{B1} = 0.982 Pv + 0.97(1.0 - Pv) + C_{B1}, \tag{5}$$

$$\varepsilon_{B2} = 0.984 Pv + 0.977(1.0 - Pv) + C_{B2}, \tag{6}$$

with the cavity terms $C_{B1}$ and $C_{B2}$ both set to 0.005. B1 makes reference to the thermal band at 11 µm and B2 to the band at 12 µm. The arithmetic mean of the emissivities for the two bands is taken as final emissivity:

$$\varepsilon = 0.5(\varepsilon_{B1} + \varepsilon_{B2}). \tag{7}$$

Both split-window and dual-angle algorithms use a set of regression coefficients $c_i$, i = 0, .., 6, for LST retrieval, as explained in [7] and mentioned in Section 2.2.1. These coefficients are obtained from simulated data using 61 atmospheric profiles representative of different atmospheric conditions. The values of these coefficients are shown in Table 3.

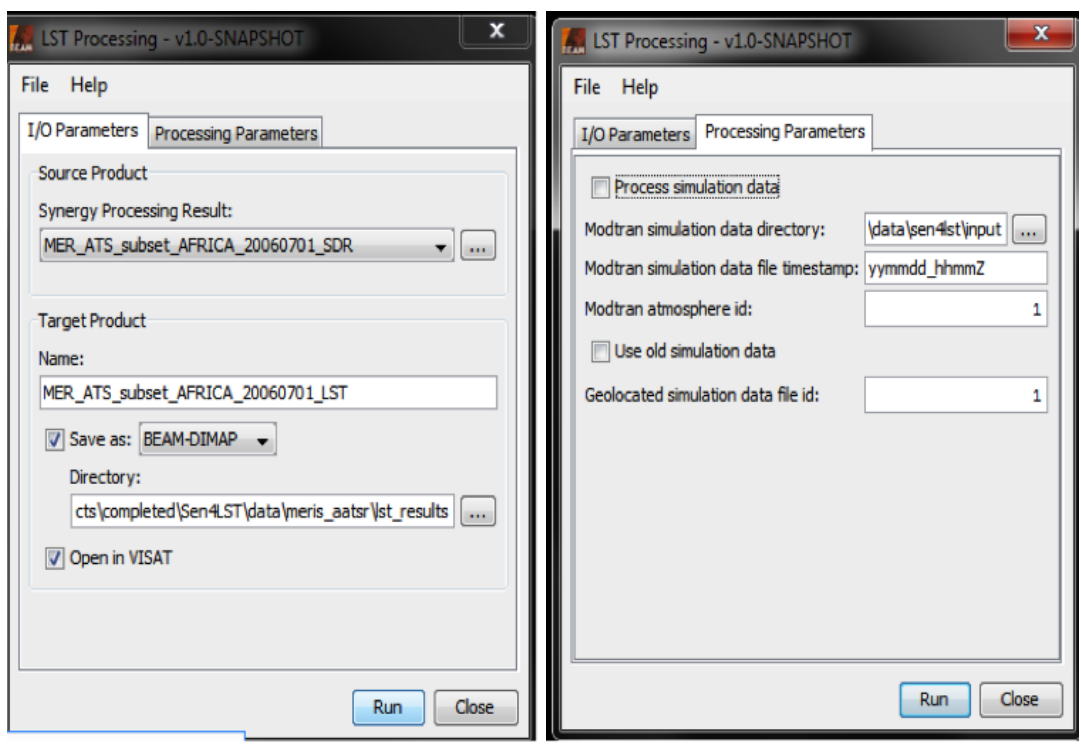

**Figure 5.** Screen-shot of the LST retrieval module in the BEAM LST plug-in.

**Table 3.** Algorithm coefficients for the DA and the SW approaches for AATSR.

| Algorithm | Coefficient Name | Coefficient Value |
|---|---|---|
| Dual-Angle (11 µm) | $C_0$ | −0.441 |
| | $C_1$ | 1.790 |
| | $C_2$ | 0.221 |
| | $C_3$ | 64.26 |
| | $C_4$ | −7.60 |
| | $C_5$ | −30.18 |
| | $C_6$ | 3.14 |
| Split-Window (11 & 12 µm) | $C_0$ | −0.268 |
| | $C_1$ | 1.084 |
| | $C_2$ | 0.277 |
| | $C_3$ | 45.11 |
| | $C_4$ | −0.73 |
| | $C_5$ | −125.0 |
| | $C_6$ | 16.70 |

The OLCI/SLSTR Simulation Mode

The SEN4LST processor provides as a second synergetic approach the combination of the OLCI and SLSTR instruments on-board Sentinel-3. Although this option must be regarded as experimental and has space for improvements, it allows a full LST retrieval from OLCI/SLSTR surface reflectance simulation datasets. With respect to the project frame, this is an "add-on" feature of the processor, but this extension is regarded as very useful toward possible Sentinel-3 applications in the future. The geolocated simulation dataset tested consisted of OLCI and SLSTR products given in ENVI format. This dataset was generated from SEN3EXP input data ([9,29]). There are separate files for OLCI (radiances, 300 m resolution), SLSTR (radiances and brightness temperatures, nadir and oblique, 500 m and 1 km resolution). For each of these products, a corresponding "coordinated" product is available, providing the geo-information as latitude/longitude bands. Four sub-scenes were simulated with a different geometry, indicated by a file suffix. There was a second type of simulation input data, and the products are given in ENVI format and are based on MODTRAN simulations with SEN3EXP input data being resampled to 500 m [30]. The products provide SLSTR nadir radiances and nadir/oblique brightness temperatures from given LST and emissivity. For more detail on the contents of these products, refer to [31]; and for the validation results, please refer to [32].

2.2.3. Application to Sample Images

As a demonstration of the potential of the SEN4LST product for climate studies, a monthly LST composite over the Iberian Peninsula for July 2009 is shown in Figure 6. The LST is generated with the SW algorithm using MERIS/AATSR imagery. This figure clearly shows the different LST patterns over Northern and Southern Spain, with lower temperatures in the North and higher temperatures in the South.

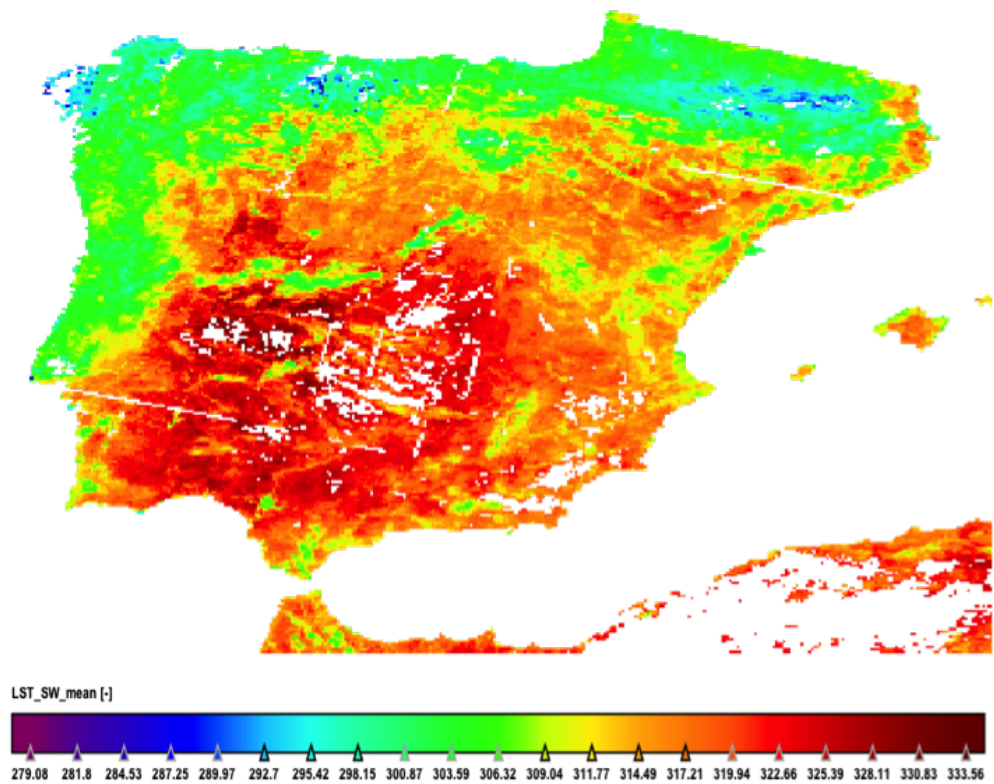

**Figure 6.** Monthly LST for July 2009 over the Iberian Peninsula.

Figure 7 shows an image showing the LST (derived from the SW algorithm) for all of Europe over a period of one month (July 2003). The spatial resolution is 3 km. The colour bar and histogram are in the lower left corner. During that summer month, many cloud-free samples for almost of Europe were available, except for Ireland and a few small areas over Eastern Europe.

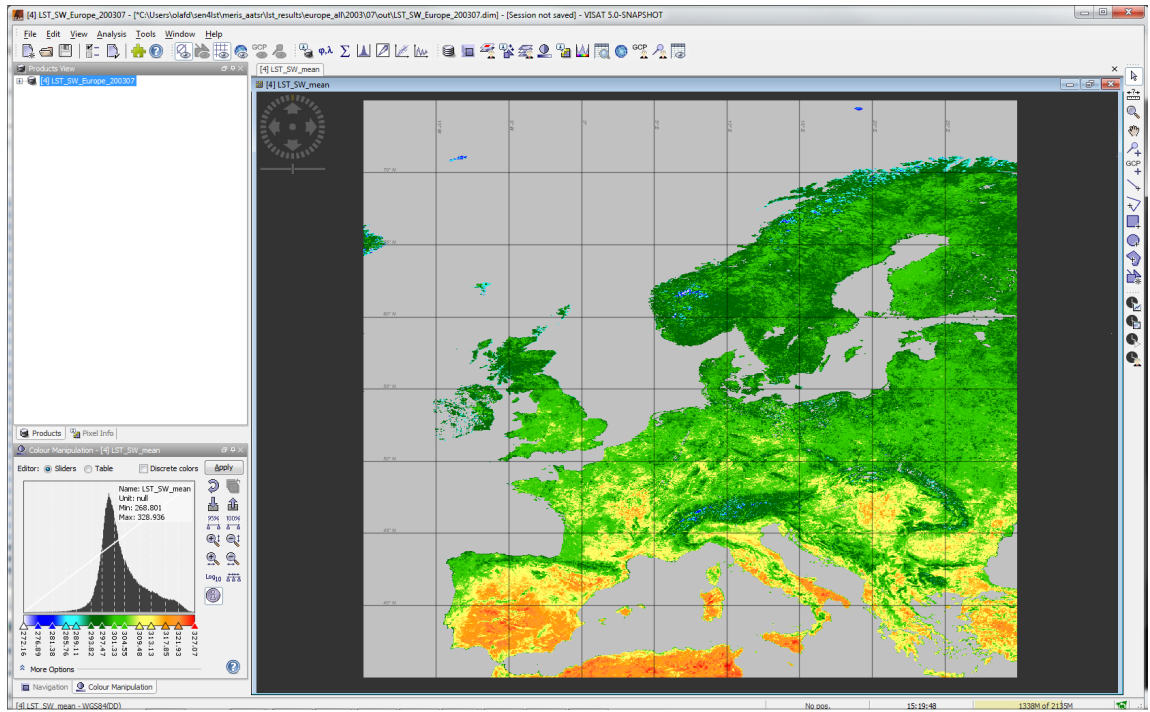

**Figure 7.** Monthly LST for July 2003 over the Europe, screen-shot of VISAT.

## 3. Discussion

The main objective of the SEN4LST project has been to exploit the synergy between S3/OLCI and S3/SLSTR data to improve LST retrievals by proposing new methodologies, but also taking advantage of the heritage of the AATSR LST standard product and the combination of MERIS/AATSR (Synergy Project). Despite several candidate methodologies having been identified, it has been concluded that the split-window method stands out for its simplicity and good performance and it has been selected to be part of the processor. The SW has the input emissivity estimated from a simple approach based on the NDVI (NDVI-THM). This algorithm provided better results in the validation made using simulated and in situ data than the standard AATSR L2 product ([7,32]). A complete database of in situ measurements, satellite imagery and simulated data was compiled. The SEN4LST Processor was also developed and implemented in BEAM, which allows the processing of the simulated data and the processing of MERIS/AATSR imagery to derive the new LST products. The results of the project have demonstrated that the LST candidate algorithm developed in the framework of the SEN4LST project can be applied in an operational processing chain to generate LST products, with potential improvements to the existing AATSR Operational product. A second synergistic approach in the processor allows for the combination of the OLCI and SLSTR instruments on-board Sentinel-3. Although this option must be regarded as experimental and has room for improvements, it allows a full LST retrieval from OLCI/SLSTR surface reflectance simulation datasets. One point under discussion in the scientific community that has not been faced here is the angular dependence of the proposed algorithm from the observation zenith angle. This is a complex matter over land and so far there is not an operational solution. Some solutions like the one in [33] are still to be quantified in absolute and relative terms. Maybe to follow some of the normalisation approaches made over ocean (BRDF) could be the path to follow.

*3.1. Potential Improvements and Updates to the Algorithm*

During the processing of the SEN4LST product and analysis of results, some potential improvements to the algorithm have been identified:

1. SW coefficients were obtained from simulated data using 61 atmospheric profiles representative of different atmospheric conditions. A new more completed atmospheric database has been recently constructed with around 5000 atmospheric profiles, which could be used to compute a new set of algorithm coefficients. It would be also possible to separate between daytime and nighttime atmospheric profiles, thus providing a set of SW coefficients for daytime acquisitions and another set of SW coefficients for nighttime acquisitions.
2. Extension of the SEN4LST product to nighttime acquisitions: the SEN4LST product was generated only for daytime acquisitions because the inputs to the algorithm, emissivity and water vapour, are generated from visible and near-infrared (VNIR) data acquired by the MERIS sensor (and VNIR data is only available for daytime acquisitions). SEN4LST product for nighttime acquisitions could be generated by using the same emissivity maps obtained from the daytime acquisitions, since it could be a good approximation. In the case of the atmospheric water vapour, whose variability is higher during the daily cycle, it would be preferable to use some external auxiliary data (i.e., re-analysis). LST methods based on day/night pairs, using the 3.7 µm of SLSTR observations could also be taken into account in future research.
3. Synergy between high resolution and low resolution data for surface emissivity retrieval: in the framework of the SEN4LST project a synergy between S2/MSI and S3/OLCI+SLSTR was proposed. However, during the realization of the project it was accorded that "operational and near-real-time processing" should be a main driver to select the best candidate algorithm [34]. For this reason, synergy was focused on OLCI and SLSTR, since both sensors are on board the same platform.

*3.2. Potential Improvements and Updates to the Processor*

The main action concerning the potential improvement of the processor would be its transference to the SNAP platform and its inclusion as a plug-in in the Sentinel-3 Toolbox. This could be accompanied by its total adaptation to Sentinel-3 OLCI and SLSTR instruments and the possibility of using it in an operational mode. In terms of adaptability, it would be convenient to be able to use the nighttime SLSTR data, merged with daylight visible data for the NDVI retrieval. This simple add would extend the analysis of the results to a complete daily cycle and make the observations more accurate and coherent.

**Supplementary Materials:** Access to the processor and the OLIC/SLTSR simulation can be provided by the authors if there is an expression of interest.

**Acknowledgments:** This work was supported by the European Space Agency (SEN4LST, ITT AO/1-6564/10/I-AM).

**Author Contributions:** Olaf Danne implemented the SEN4LST algorithm in Java to be included in a BEAM module. Norman Fomferra is the founder of BEAM, together with Carsten Brockmann who is also the CEO of Brockmann Consult. Ana Ruescas contributed to the development of the SEN4LST project and wrote the paper. All authors have read and approved the final manuscript.

**Conflicts of Interest:** The authors declare no conflict of interest.